\documentstyle[12pt,epsf]{article} 
\newcommand{\lsi}{\raise0.3ex\hbox{$<$\kern-0.75em\raise-1.1ex\hbox{$\sim$}}}
\newcommand{\gsi}{\raise0.3ex\hbox{$>$\kern-0.75em\raise-1.1ex\hbox{$\sim$}}}

\newcommand{\gsim}{\mathop{\gsi}}
\renewcommand{\vec}[1]{{\bf #1}}

\newcommand{\rhs}{r.h.s.\ }
\newcommand{\lhs}{l.h.s.\ }

\renewcommand{\(}{\left(}
\renewcommand{\)}{\right)}
\renewcommand{\[}{\left[}
\renewcommand{\]}{\right]}

\newcommand{\mmdebye}{m^2_{\rm D}}

\newcommand{\tr}{{\rm t}}

\newcommand{\vA}{\vec{A}}
\newcommand{\vE}{\vec{E}}
\newcommand{\va}{\vec{a}}
\newcommand{\ve}{\vec{e}}
\newcommand{\vk}{\vec{k}}
\newcommand{\vp}{\vec{p}}
\newcommand{\vv}{\vec{v}}
\newcommand{\vx}{\vec{x}}

\newcommand{\intv}[1]{\int\frac{d\Omega_{{\rm v}_{#1}}}{4\pi}}
\newcommand{\deltav}{\delta^{(S^2)}}
\newcommand{\mlabel}[1]{\label{#1}}
\newcommand{\eq}[1]{Eq.~\mref{#1}}
\newcommand{\eqs}[1]{Eqs.~\mref{#1}}
\newcommand{\mref}[1]{(\ref{#1})}
\newcommand{\rfs}[1]{ Refs.~\cite{#1}}
\newcommand{\rf}[1]{ Ref.~\cite{#1}}

\begin{document}
 
\setlength{\baselineskip}{0.6cm}
\newcommand{\nn}{\nonumber}
\newcommand{\figysize}{16.0cm}
\newcommand{\figtopspace}{\vspace*{-1.5cm}}
\newcommand{\figbottomspace}{\vspace*{-5.0cm}}

\begin{titlepage}
\begin{flushright}
HD-THEP-98-04\\
January 1998
\end{flushright}
\begin{centering}
\vfill
 
{\bf Effective dynamics of soft non-Abelian gauge fields at
finite temperature }
\vspace{1cm}

Dietrich B\"odeker\footnote{bodeker@thphys.uni-heidelberg.de} 

\vspace{1cm} {\em 
Institut f\"ur Theoretische Physik, Universit\"at Heidelberg, 
Philosophenweg 16, 
D-69120~Heidelberg, Germany}

\vspace{2cm}
 
{\bf Abstract}

\vspace{0.5cm}

We consider time dependent correlation functions of non-Abelian gauge
fields at finite temperature. An effective theory for the soft ($p\sim
g^2 T$) field modes is derived by integrating out the field modes with
momenta of order $T$ and of order $g T$ in a leading logarithmic
approximation.  In this effective theory the time evolution of the
soft fields is determined by a local Langevin-type equation. As an
application, the rate for hot electroweak baryon number violation is
estimated as $\Gamma \sim g^2 \log(1/g) (g^2 T)^4$. Furthermore,
possible consequences for non-perturbative lattice computations of
unequal time correlation functions are discussed.
\end{centering}

\vspace{0.5cm}\noindent


\vspace{0.3cm}\noindent
 
\vfill \vfill
\noindent
 
\end{titlepage}
 

Finite temperature field theory for non-Abelian gauge fields is
non-perturbative for soft\footnote{For spatial vectors we use the
  notation $k=|\vk|$. 4-momentum vectors are denoted by $K^\mu =
  (k^0,\vk)$ and we use the metric $K^2 = k_0^2 - k^2$. $P$ always
  refers to a soft momentum and $K$ to a semi-hard or a hard one.}
momenta $p\sim g^2 T$, where $g$ is the gauge coupling and $T$ is the
temperature ~\cite{linde80,gross}. The dynamics of the soft gauge
fields determines the rate for electroweak baryon number violation at
very high temperatures \cite{rubakov}. This rate is determined by a
real time correlation function of the type
\begin{eqnarray}
   C(t_1 - t_2) = \langle 
   {\cal O}[A(t_1)]  {\cal O} [A(t_2)]\rangle
        \mlabel{c}
\end{eqnarray} 
where $\langle \cdots \rangle$ denotes the thermal average and ${\cal
O}[A(t)]$ is a gauge invariant function of the gauge fields $A_\mu^a(t,\vx)$.

The aim of this letter is to derive an effective theory for the
dynamics of the soft gauge fields by integrating out the ``hard''
($k\sim T$) and ``semi-hard'' ($k\sim gT$) fields. For weak
gauge coupling $g$ this can be done perturbatively. This effective
theory should allow for a non-perturbative computation of real time
correlation functions like \mref{c}, e.g., on a lattice. 

For time-independent physical quantities, like the free energy or
correlation lengths, the effective theory for soft gauge fields is
well established. It is a three-dimensional gauge theory, the
parameters of which are determined by dimensional reduction
\cite{farakos,braateneffective}. The dimensionally reduced theory can
serve as an input for Euclidean lattice simulations \cite{lattice} or
other non-perturbative methods \cite{gap}. The three-dimensional
theory is much simpler to simulate on a lattice than the full
four-dimensional one.

For time dependent quantities it may be even more important to find an
effective theory for the soft field modes: There is no apparent way to
calculate real time quantities like \mref{c} in lattice simulations of
finite temperature quantum field theory. Presently the only known tool
to evaluate them is the classical field approximation
\cite{grigoriev}-\cite{ambjorn} and variants thereof
\cite{bodeker95}-\cite{iancu} which contain additional
degrees of freedom representing the hard field modes.

The reason why the classical field approximation is expected to be
reliable is that the field modes with soft momenta have a large
occupation number and should therefore behave classically. The high
momentum modes have occupation number of order unity and do not behave
like classical {\em fields}.  They are however weakly interacting and
can be treated as almost free massless {\em particles}
moving quasi-classically.

There has been a long discussion (see Refs.\ 
\cite{mclerran,khlebnikov}) about whether these hard particles do
affect the dynamics of the soft field modes, and, in particular,
whether they play a role in electroweak baryon number violation. If
all field modes with momenta larger than $g^2 T$ were irrelevant,
there would be only one scale in the problem and on dimensional
grounds the rate can be estimated as $\Gamma\sim (g^2T)^4$
\cite{khlebnikov}. Arnold {\em et al.} \cite{asy} demonstrated that
the dynamics of the soft fields is damped by the hard particles and
they obtained the estimate $\Gamma\sim g^2 (g^2T)^4$.  There are,
however, contributions to the soft dynamics due to semi-hard field
modes which are as important as the hard particles \cite{bodeker95}.
These contributions are the subject of this letter. Here only the main
points of the calculation will be described. Further details can be
found in \rf{preparation}.

The first step in deriving an effective theory for soft gauge fields
is to integrate out the hard field modes with spatial momenta of order
$T$. The dominant contributions from these fields are the so called
hard thermal loops \cite{pisarski}. A consistent perturbative
expansion for the semi-hard field modes requires the use of the so
called Braaten-Pisarski scheme, or hard thermal loop effective
theory\footnote{For certain momenta $K= (k_0,\vk)$ with $k$ of order
  $g T$, however, the hard thermal loop effective expansion breaks
  down, e.g., when $K$ is on the light cone \cite{flechsig}.}.  In
this scheme the hard thermal loop propagators and vertices must be
treated on the same footing as their tree-level counterparts.

In the present context integrating out the hard field modes yields an
effective theory for the semi-hard and the soft fields. As we will see
below, the leading contributions to the effective soft dynamics arise
from hard thermal loop induced interactions of the soft fields with
the semi-hard ones. Since there are no hard thermal loop vertices in
Abelian theories, the terms we are going to compute do not have an
Abelian analogue (cf.\ \rfs{lebedev,carrington}).

In order to integrate out the semi-hard fields 
we introduce a separation scale $\mu$ such that 
\begin{eqnarray}
        g^2 T\ll \mu \ll gT
\end{eqnarray}
and integrate out all field modes with $k>\mu$.

\begin{figure}[tb]
 
\vspace*{-4.0cm}
 
\hspace{1cm}
\epsfysize=25cm
\centerline{\space{5cm}\epsffile{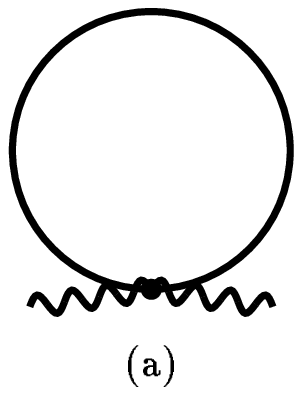}
        \hspace{-13cm}\epsfysize=25cm\epsffile{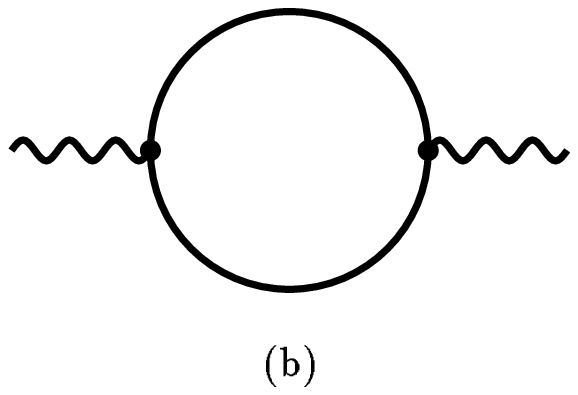}
        \hspace{-12cm}\epsfysize=25cm\epsffile{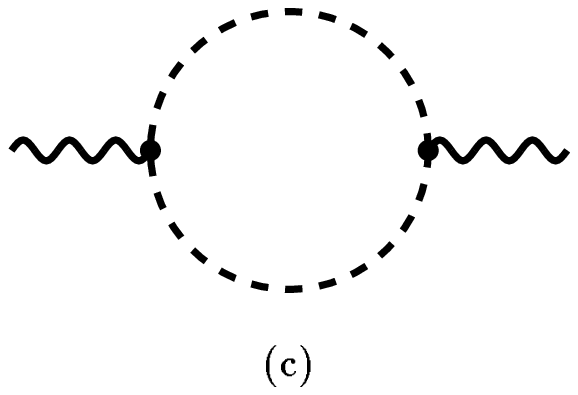}}

\vspace*{-18cm}
 
\caption[a]{Sub-leading contributions to the effective theory for the
soft ($p\sim g^2T$) gauge fields. The loop momenta are semi-hard
($k\sim gT$).  The small dots are ``bare'' (non hard thermal loop)
vertices. The full lines denote hard thermal loop resummed
propagators, the dashed lines represent ghost propagators.}
\label{treevertex}
\end{figure}
Let us first discuss what kind of diagrams are relevant to our
problem.  We have to consider diagrams in the hard thermal loop
effective theory in which all internal loop momenta are semi-hard and
the external momenta are soft. These have to be compared with the
corresponding ``tree-level'' terms like, i.e., the transverse hard
thermal loop
polarization operator which is given by
\begin{eqnarray}
        \delta \Pi_{\rm t}(P) = \frac12 P^{ij}_{\rm t}(\vp) 
        \, \, \mmdebye \intv{} v^i v^j \frac{p^0}{v\cdot P}.
        \mlabel{polarization}
\end{eqnarray}
Here $P^{ij}_{\rm t}(\vp)$ is the transverse 
projector 
\begin{eqnarray}
  P^{ij}_{\rm t}(\vp) = \delta^{ij} - \frac{p^i p^j}{p^2} 
\end{eqnarray}
and $\mmdebye$ is the leading order Debye mass squared. In a pure
SU($N$) gauge theory, it is given by $\mmdebye = (1/3) N g^2 T^2$. In
the hot electroweak theory it receives additional contributions due to
the Higgs and fermion fields. Furthermore, $v^\mu \equiv
(1,\vec{v})$, and the integral $\int d\Omega_{\vv}$ is over the
directions of the unit vector $\vv$, $|\vv| =1$ .

Consider the corrections to \eq{polarization} due to the semi-hard
field modes. The leading order behavior of the diagrams in Fig.\
\ref{treevertex} can be easily estimated since they are completely
analogous to hard thermal loop diagrams for semi-hard external momenta
except that now all momenta are smaller by a factor $g$.
The $k$ integral behaves like $\int dk\sim gT $ which gives a
suppression factor $g$ relative to the hard thermal loops.  In other
words, the diagrams in Fig.\ \ref{treevertex} can be neglected
relative to the ``tree-level'' term \mref{polarization}. 

\begin{figure}[t]
 
\vspace*{-4.0cm}
 
\hspace{1cm}
\epsfysize=25cm
\centerline{\space{5cm}\epsffile{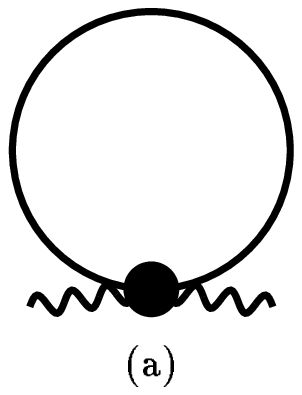}
        \hspace{-11cm}\epsfysize=25cm\epsffile{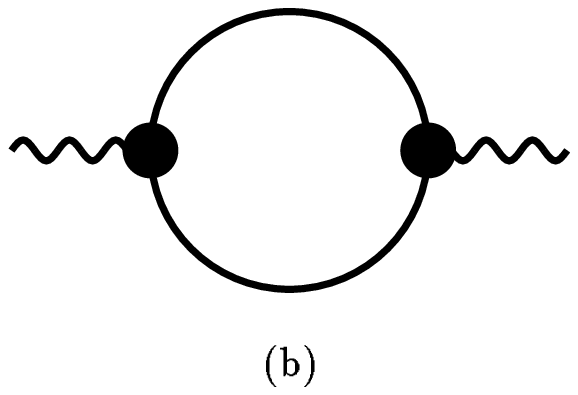}
        }

\vspace*{-18cm}
 
\caption[a]{Leading contributions to the effective theory for the
soft ($p\sim g^2T$) gauge fields. The loop momenta are semi-hard
($k\sim gT$).  The full lines denote hard thermal loop resummed
propagators and the heavy dots  are hard thermal loop vertices.
}
\label{htlvertex}
\end{figure}
Now consider the analogous diagrams (Fig.\ \ref{htlvertex}) with the
``bare'' vertices replaced by hard thermal loop vertices.  There are
only two such diagrams since there is no hard thermal loop vertex
involving ghost fields. The vertex in Fig.\ \ref{htlvertex}(a)
contains a term proportional to $g^2 p_0/(v\cdot P)^2$ times a factor
of order $gT$.  Here the four vector $v^\mu$ is of the same type as in
\eq{polarization}. Due to the Bose distribution function in the loop
integral there is a factor $T$ so that diagram 2(a) can be estimated
as $\mmdebye g^2 T p_0/(v\cdot P)^2$.  Comparing this expression with
\eq{polarization} we see that both terms are of the same order of
magnitude. In fact, the diagrams in Fig.\ \ref{htlvertex} can be even
larger by a factor of $\log(gT/\mu)$: The transverse propagator is
unscreened when $|k_0|\ll k$ and in the infrared the loop integrals
can be sensitive to the separation scale $\mu$.  We will see below
that this logarithm indeed occurs. It is specific to the transverse
semi-hard field modes. In this letter we will compute only these
leading logarithmic contributions to the effective soft dynamics.

\begin{figure}[tb]
 
\vspace*{-4.0cm}
 
\hspace{1cm}
\epsfysize=25cm
\centerline{\epsffile{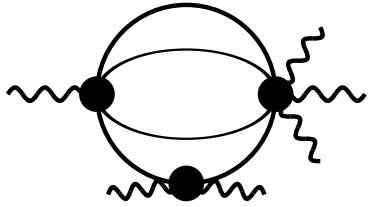}}

\vspace*{-18cm}
 
\caption[a]{Contribution to the effective theory for the soft ($p\sim
g^2T$) gauge fields involving hard thermal loop $n$-point
vertices. The notation is the same as in Fig.\ \ref{htlvertex}.}
\label{many}
\end{figure}
In order to obtain the effective theory for the soft field modes we
have to consider diagrams with more external soft fields and with more
internal semi-hard propagators like in Fig.\ \ref{many}.  Higher hard
thermal loop $n$-point vertices contain a large number of terms which make a
diagrammatic analysis very difficult. It is more convenient to use an
effective field theory description and solve the corresponding field
equations of motion.

There are local\footnote{The hard thermal loop vertices are non-local in 
space and time.} formulations of the hard thermal loop effective theory
due to Blaizot and Iancu \cite{blaizot93} and due to Nair \cite{nair}.
These formulations make the physical content of the hard thermal
loop effective theory quite transparent (see also \rf{kelly}). The
formulation due to Blaizot and Iancu is the non-Abelian generalization
of the Vlasov equations for a QED plasma. It consists of Maxwell's
equation for the semi-hard and soft gauge fields, i.e., fields with
spatial momenta of order $gT$ or less,
\begin{eqnarray}
        \mlabel{maxwell}
        [D_\mu, F^{\mu\nu}(x)]=  j^\nu (x)
\end{eqnarray}
where $D_\mu = \partial_\mu - i g A_\mu(x)$ is the covariant
derivative and $A_\mu(x) = A_\mu^a(x)T^a$ with hermitian generators
normalized to Tr$(T^a T^b) =(1/2) \delta^{ab}$. The current $j^\nu(x)$
is due to the hard field modes. These modes are weakly interacting and
they behave as massless particles moving at the speed of light with
3-velocity $\vv$.  The current can be written as
\begin{eqnarray}
   j^\nu(x) = \mmdebye \intv{}
        v^\nu  W(x,v).
        \mlabel{current}
\end{eqnarray}
The field $W(x,v) = W^a(x,v) T^a$ is proportional to the
deviation of the distribution of hard particles from the equilibrium
distribution.  It is determined by the equation of motion
\begin{eqnarray}
        [v \cdot D, W(x,v)] = \vec{v}\cdot\vec{E} (x) ,
        \mlabel{vlasov}
\end{eqnarray}
where $\vec{E}$ is the electric field strength. The conserved
Hamiltonian corresponding to Eqs.\ \mref{maxwell}-\mref{vlasov} is
\cite{blaizot94}
\begin{eqnarray}
        \mlabel{hamiltonian} H=\int d^3 x {\rm Tr} \left\{
        \vec{E}(x)\cdot\vec{E}(x) +\vec{B}(x)\cdot\vec{B}(x) +
        \mmdebye\int\frac{d\Omega_\vec{v}}{4\pi}
        W(x,v) W(x,v)\right\}.
\end{eqnarray}

Originally, the hard thermal loop effective theory was designed for
calculating Green functions with semi-hard momenta.
Following~\cite{bodeker95} we assume that the leading order dynamics
for both soft and semi-hard fields is correctly described by the hard
thermal loop effective theory. More precisely, we assume that real
time correlation functions of gauge invariant operators like \mref{c}
can be obtained as follows:
\begin{itemize}
\item Compute the solution $A_\mu^a(x)$ of the non-Abelian Vlasov
  equations \mref{maxwell}-\mref{vlasov} for given initial conditions
  $A_{\rm in}(\vx)$, $\vec{E}_{\rm in}(\vx)$ and $W_{\rm in}(\vx,v)$
  at $t=0$.
\item Then $C(t_1 - t_2)$ is given by the product ${\cal O}[A(t_1)]
  {\cal O}[A(t_2)]$ averaged over the initial conditions with the
  Boltzmann weight $\exp(-\beta H[A_{\rm in},\vec{E}_{\rm
  in},W_{\rm in}])$ where Gauss' law has to be imposed as a
  constraint.
\end{itemize}
In the average over initial conditions ultraviolet divergences occur.
These are due to the fact that a classical field theory at finite
temperature is not well defined.  One has to use an UV cutoff
$\Lambda$ which satisfies
\begin{eqnarray}
gT\ll \Lambda \ll T . \mlabel{htlcutoff}
\end{eqnarray}
After the semi-hard field modes have been integrated out, the
dependence on $\Lambda$ cancels against the appropriate
$\Lambda$-dependent counter-terms to be included in the hard thermal
loop effective theory\footnote{See also the discussion in \rf{iancu}}.

To integrate out the semi-hard modes, we decompose  $A$, $\vec{E}$ and $W$
into
\begin{eqnarray}
   A &\to& A + a\nn \\
   \vec{E} &\to& \vec{E} + \vec{e}\nn \\
   W &\to& W + w . \mlabel{separation}
\end{eqnarray}
The soft modes\footnote{For notational simplicity we do not introduce
new symbols for the soft modes. From now on $A$, $\vec{E}$ and $W$
will always refer to the soft fields only.} $A$, $\vec{E}$ and $W$
contain the spatial Fourier components with $p<\mu$ while the
semi-hard modes $a$, $\ve$ and $w$ consist of those with $k>\mu$.  Let
us note that the separation \mref{separation} does not respect gauge
invariance. Nevertheless, the results we are going to obtain, will not
depend on the choice of a gauge. Furthermore, we do not specify the way
the cutoff is realized precisely. The only cutoff dependence which we
will find is a logarithmic one for which the precise choice should be
irrelevant. Here we consider only the transverse
semi-hard gauge fields. 

Due to the non-linear terms in \eqs{maxwell} and \mref{vlasov} the
equations of motion for the soft and the semi-hard fields are coupled.
Let us see which of these couplings are relevant to our problem. The
non-linear terms in Maxwell's equations \mref{maxwell} correspond to
non-hard thermal loop vertices. For the interaction between soft and semi-hard
fields these vertices can be neglected as we have already argued. The
important vertices are the hard thermal loop ones. They correspond to
non-linear terms in \eq{vlasov}. Thus, substituting \eq{separation}
into \eqs{maxwell}-\mref{vlasov}, the equations for the soft fields
read
\begin{eqnarray}
        [D_\mu, F^{\mu\nu}(x)]=  \mmdebye \int\frac{d\Omega_\vec{v}}{4\pi}
        v^\nu  W(x,v)
        \mlabel{maxwellsoft}
\end{eqnarray}
and
\begin{eqnarray}
        [v \cdot D, W(x,v)] = \vec{v}\cdot\vec{E} (x)
        + \xi(x,v) ,
        \mlabel{vlasovsoft}
\end{eqnarray}
where
\begin{eqnarray}
        \xi^a (x,\vec{v}) = g f^{abc} \( \vec{v}\cdot\vec{a}^b_\tr (x)
        w^c(x,v) \)_{\rm soft} .
        \mlabel{xi}
\end{eqnarray}
Here the subscript ``soft'' indicates that only spatial Fourier
components with $p<\mu$ are included. The structure constants
$f^{abc}$ are defined such that $[T^a,T^b] = i f^{abc} T^c$.  The
transverse gauge field $\vec{a}_\tr (x)$ and the field $w(x,v)$ in
\eq{xi} are to be understood as the solutions to the coupled equations
of motion \mref{maxwellsoft}-\mref{xi} and
\mref{maxwellhard}-\mref{h}.

In the following we discuss how the semi-hard fields can be eliminated
from Eqs.\ \mref{maxwellsoft}-\mref{xi} leading to the final result 
\mref{boltzmann}.  A more detailed description of the calculation can be
found in \rf{preparation}.

The semi-hard fields are weakly interacting and can therefore be
treated perturbatively.  Since we are only interested in leading order
results, it might appear sufficient to insert the solutions of the
free equations of motion, $\vec{a}_\tr {}_0 (x)$ and $w_0(x,v)$ into
$\xi(x,v)$. The corresponding term will be called $\xi_0(x,v)$.  As
will be seen below, a next-to-leading order term has to be included as
well. For this term, which will be denoted $\xi_1(x,v)$, one can
neglect those interactions in \eq{vlasov} which contain only semi-hard
fields.  In the diagrammatic language such interactions correspond to
parts of hard thermal loop vertices which do not contain the enhancing
factors $1/(v\cdot P)$ discussed above.  The relevant non-linear terms
are the ones which contain a product of a soft and a semi-hard field.
Thus the equations of motion in spatial momentum space can be written
as
\begin{eqnarray}
        \mlabel{maxwellhard}
        \ddot{\va}^a_\tr(t,\vk) + k^2 \va^a_\tr (t,\vk) =
         \mmdebye \int\frac{d\Omega_v}{4\pi}
        \vv_{\tr} w^a(t,\vk,v) ,
\end{eqnarray}
\begin{eqnarray}
        \mlabel{vlasovhard}
        \dot{w}^a (t,\vk,v) + i \vv\cdot\vk w^a(t,\vk,v) =
        \vv\cdot \ve_\tr  (t,\vk) +   h^a (t,\vk,v) ,
\end{eqnarray}
where $v^i_{\tr} = P^{ij}_{\rm t}(\vk) v^j $ is the transverse
projection of $\vv$. The term $ h^a (t,\vk,v)$ contains the interaction of the
semi-hard fields with the soft ones. In coordinate space it reads
\begin{eqnarray}
         h^a (x,v) = 
        g f^{abc}\( -v\cdot A^b(x) w^c(x,v)
                   +\vv\cdot\va^b_\tr (x) W^c(x,v)\).
        \mlabel{h}
\end{eqnarray}

A convenient method for solving Eqs.\ \mref{maxwellhard}-\mref{h} is
the Laplace transformation\footnote{Or one sided Fourier
  transformation, see,e.g., Ref.~\cite{landau10}.}. The Laplace
transform of a function $f(t)$ is defined as
\begin{eqnarray}
       f(k^0) \equiv \int_0^\infty dt e^{i k^0 t} f(t)
        \mlabel{laplace}
\end{eqnarray}
and it is an analytic function in the upper half of the complex
$k^0$-plane.  Applying \mref{laplace} to  \eqs{maxwellhard} and
\mref{vlasovhard} one obtains
\begin{eqnarray}
        a^{ia}_\tr(K) =
         a^{ia}_{\tr 0}(K) + \intv{1} \Delta_{12}^i(K,v_1) h^a(K,v_1)
        \mlabel{solutiona}
\end{eqnarray}
and
\begin{eqnarray}
        w^{a}(K,v) = w^{a}_0 (K,v) + \intv{1} \Delta_{22}^i(K,v,v_1) h^a(K,v_1)
        , \mlabel{solutionw}
\end{eqnarray}
where $\vec{a}_\tr {}_0(K)$ and $w_0(K,v)$ are the solutions to the free
equations of motion. They are determined by the initial values at $t=0$.
The propagator functions in \mref{solutiona}, \mref{solutionw} are given by
\begin{eqnarray}
        \Delta_{12}^{i}(K,v) &=& i \mmdebye v^j G^{ij}_{\rm t}(K)      
        \frac{1}{v\cdot K} , \\
        \Delta_{22}(K,v,v_1) &=& 
        4\pi \deltav(\vv - \vv_1) \frac{i}{v\cdot K} - 
        \mmdebye v^i v_1^j G^{ij}_{\rm t}(K)    
        \frac{i k^0}{v\cdot K v_1\cdot K} .
\end{eqnarray}
Here $G^{ij}_{\rm t}(K)$ is the transverse hard thermal loop resummed
propagator,
\begin{eqnarray}
        G^{ij}_{\rm t}(K) =     
          \frac{1}{-K^2 + \delta \Pi_{\rm t}(K)}
         P^{ij}_{\rm t}(\vk)  
        \mlabel{propagator} .
\end{eqnarray}
Furthermore, $\deltav$ is the delta function on the two dimensional unit
sphere:
\begin{eqnarray}
        \intv{1} f(\vv_1) \deltav(\vv - \vv_1) = f(\vv) .
\end{eqnarray}

The solutions to the equations of motion, Eqs.~\mref{solutiona} and
\mref{solutionw}, are still formal in the sense that $h^a(K,v)$
contains the complete solutions $\va_\tr(K)$ and $w(K,v)$ as well as
$A(P)$ and $W(P,v)$. However, by iterating Eqs.~\mref{solutiona} and
\mref{solutionw} one obtains a series in which each term contains only
the free solutions $\va_{\tr 0}(K)$ and $w_0(K,v)$ together with the
full $A(P)$ and $W(P,v)$.  Since \eqs{maxwellhard} and
\mref{vlasovhard} are linear in $\va_\tr$ and $w$, the solutions are
linear in $\va_{\tr 0}$ and $w_0$.  The function $\xi(x,v)$ is
therefore bilinear in the fields $\va_{\tr 0}$ and $w_0$ which always
appear in the form
\begin{eqnarray}
        \chi = \phi(K-P) \phi'(P' -K) ,
        \mlabel{chi}
\end{eqnarray}
where the $\phi$ and $\phi'$ are either $\va_{\tr 0}$ or $w_0$. The solution
to the (non-perturbative) equations of motion for the soft fields will
contain products of $\chi$'s.  To obtain a correlation function like
\mref{c} one has to average over the initial conditions for both soft
and semi-hard fields.

In general, the thermal average of a product $\chi_1\cdots \chi_n$ can
be approximated by disconnected parts: The connected part $\langle
\chi_1\cdots \chi_n\rangle_{\rm c}$ has one momentum conserving delta
function for a soft momentum. A disconnected part like for instance
$\langle \chi_1\cdots \chi_m\rangle_{\rm c}\langle \chi_{m+1}\cdots
\chi_n\rangle_{\rm c}$ contains two such delta functions. The
corresponding contribution to a non-perturbative correlation function
has one soft 4-momentum integration less and one additional semi-hard
4-momentum integration. We can therefore estimate\footnote{As we have
already discussed, the integration over semi-hard momenta can give
logarithms of the separation scale $\mu$. These logarithms are not
included in this estimate.}
\begin{eqnarray}
  \Big\langle\chi_1\cdots
  \chi_n
    \Big\rangle_{\rm c}  \sim  g^3 \frac{p^0}{gT}
    \Big\langle\chi_1 \cdots\chi_m
    \Big\rangle_{\rm c}
    \Big\langle\chi_{m+1} \cdots\chi_n
    \Big\rangle_{\rm c} .
\end{eqnarray}
Here $p^0$ is the typical frequency scale for the soft
non-perturbative dynamics. This scale will be estimated from the
equations of motion for the soft fields which we are going to derive
(see \eq{p0}).  This consideration suggests that $\xi(x,v)$ in
\eq{vlasovsoft} can be simplified by replacing the $\chi_i$ by their
thermal averages: $ \chi_i \to \langle\chi_i\rangle $. For the lowest
order term $\xi_0$, which does not depend on the soft fields, this
replacement gives zero due to the antisymmetry of the structure
constants $f^{abc}$. Therefore we have to leave $\xi_0$ in
Eq.~\mref{vlasovsoft} as it stands and the solution to the equations
of motion for the soft fields will contain products of $\xi_0$'s.  The
thermal averages of these products can be approximated by a product of
two point correlators. We also have to include the next-to-leading
order term $\xi_1$ which is linear in the soft fields $A$ and $W$.

Neglecting the external soft 4-momentum inside loop integrals
the leading logarithmic result for the two point
correlator of $\xi_0$ in configuration space reads 
\begin{eqnarray}
        \Big\langle 
    \xi_0^{a}(x_1,v_1)
    \xi_0^{b}(x_{2},v_{2}) 
    \Big\rangle 
    =
    -2 N \frac{g^2 T^2}{\mmdebye}
        \log\(\frac{g T}{\mu}\)
         I(v_1,v_2)
        \delta^{ab} \delta^{(4)}(x_1 - x_2),
        \mlabel{xi0correlator4}
\end{eqnarray}
with
\begin{eqnarray}
  \mlabel{k}
  I(v,v_1) \equiv -\deltav(\vv - \vv_1) 
        + \frac{1}{\pi^2}
        \frac{(\vv\cdot\vv_1)^2}{\sqrt{1 - (\vv\cdot\vv_1)^2}} .
\end{eqnarray}
The logarithmic $\mu$-dependence in \eq{xi0correlator4} arises because
the transverse propagator \mref{propagator} is unscreened for $|k^0|
\ll k$.  Non-logarithmic contributions have been neglected in
\eq{xi0correlator4}. To compute them, one would have to take into
account the longitudinal semi-hard gauge fields as well.  Since the
external soft momentum has been neglected in order to obtain
\eq{xi0correlator4}, this expression is valid only when applied to
problems with a relevant time scale much larger than
$(gT)^{-1}$. 

For $\xi_1$ we can replace the products of the free semi-hard fields
$\va_{\tr 0}$ and $w_0$ by their expectation values. Then the terms
containing $A(x)$ cancel. With the same approximations which were used
for \eq{xi0correlator4} one obtains
\begin{eqnarray}
   \xi_1(x,v) = N g^2 T  \log\(\frac{gT}{\mu}\)
 \intv{1}         
   I(v,v_1)  W(x,v_1).
  \mlabel{xi1replace3}
\end{eqnarray}
Note that without the cancelation of the $A(x)$-dependent terms
\eq{boltzmann} would not be gauge covariant.

With the above approximations the equation of motion \mref{vlasovsoft}
reads
\begin{eqnarray}
        [v \cdot D, W(x,v)] &=& \vec{v}\cdot\vec{E} (x)
        + \xi_0(x,v) + N g^2 T  \log\(\frac{g T}{\mu}\)
 \intv{1}   I(v,v_1)      
 W(x,v_1) . \nn\\ &&
        \mlabel{boltzmann}
\end{eqnarray}
Let us see whether Eq.~\mref{boltzmann} is consistent with Maxwell's
equation \mref{maxwellsoft} for the soft fields which requires the
current on the \rhs of \eq{maxwellsoft} to be covariantly conserved.
Integrating Eq.~\mref{boltzmann} over the direction of $\vv$ the third
term on the \rhs drops out due to
\begin{eqnarray}
        \intv{} I(v,v_1) = 0 \mlabel{vanish}
\end{eqnarray}
and one obtains
\begin{eqnarray}
        [D_\mu,j^\mu(x)] = \mmdebye \intv{} \xi_0(x,v) .
\mlabel{nonconservation}
\end{eqnarray}
That is, the current appears not to be conserved. However, as we have
argued, only the two point function of $\xi_0$ should be relevant
for the leading order behavior of a soft correlator. But the two
point function of $\xi_0(x_1,v_1)$ with the \rhs of
\eq{nonconservation} vanishes due to \eq{vanish}.  Thus we can
replace
\begin{eqnarray}
        \intv{} \xi_0(x,v) \to 0
        \mlabel{intxi0}
\end{eqnarray}
so that the current is covariantly conserved within the present
approximation.

Let us further note that, due to the term $\xi_0$, Eq.~\mref{boltzmann}
appears to be non-covariant under gauge transformations.
Nevertheless, a gauge invariant correlation function computed via
Eqs.\ \mref{xi0correlator4} and \mref{boltzmann} will not depend on
the choice of the gauge after the thermal average over initial
conditions since the two point function
\mref{xi0correlator4} is invariant under gauge transformations of
$\xi_0$.

Eq.~\mref{boltzmann} is a Boltzmann equation for the soft fluctuations
of the particle distribution $W(x,v)$. The \rhs contains a collision
term which is due to the interactions with the semi-hard fields. When
$\partial_i W $ on the \lhs of \eq{boltzmann} is of order $ (g^2 T)
W$, the collision term is larger than $\partial_i W $ by a factor
$\log(gT/\mu)$.  The reason is that the mean free path for the
hard particles interacting with fields with $k>\mu$ is of order $(g^2
T \log(gT/\mu))^{-1}$ which is smaller than the soft length scale
$(g^2 T)^{-1}$ by the same logarithm.  The collision term is
accompanied by the noise term $\xi_0$ which is due to the thermal
fluctuation of initial conditions\footnote{The term $\xi_0(x,v)$
  should not be confused with the stochastic force discussed in
  Refs.~\cite{asy,huet,arnold}. The latter is due to the fluctuations
  of the initial conditions for the soft $W(x,v)$ and it is also
  present in the theory described by Eq.\mref{boltzmann}.} of the
fields with $k>\mu$.

For a QED plasma there is no collision term at this order in the
coupling constant. In this case the size of the collision term is
determined by the transport cross section which corresponds to a mean free
path of order order $(e^4 T)^{-1}$ (cf. the discussion in \rf{huet}).
For a non-Abelian plasma the relevant mean free path is determined by
the total cross section which is dominated by small angle scattering:
Even a scattering process which hardly changes the momentum of a hard
particle can change its color charge which is what is seen by the soft
gauge fields.

What can one learn from \eq{boltzmann} about the time scale relevant
to non-perturbative physics? Let us first see how many field modes can
be integrated out perturbatively. The mean free path for the hard
particles interacting with the modes with $k>\mu$ is of order $(g^2 T
\log(gT/\mu))^{-1}$. Perturbation theory for the hard particles must
break down at this length scale. Thus we can decrease $\mu$ until $\mu
\sim g^2 T \log(gT/\mu)$. Then the logarithm in \eq{boltzmann}, in a
first approximation, becomes $\log(1/g)$. 

We will now simplify \eq{boltzmann} by neglecting terms which are
suppressed by inverse powers of $\log(1/g)$. We introduce moments of
$W(x,v)$ and $\xi_0(x,v)$,
\begin{eqnarray}
        W (x) &\equiv& \intv{} W(x,v),\\
        W^{i_1\cdots i_n} (x) &\equiv& \intv{} v^{i_1} \cdots  v^{i_n} W(x,v)
        ,\\
        \xi_0^{i_1\cdots i_n} (x) &\equiv& \intv{} v^{i_1} \cdots  v^{i_n} 
        \xi_0(x,v) .
\end{eqnarray} 
Multiplying Eq.~\mref{boltzmann} with the appropriate factors $
v^{i_1} \cdots v^{i_n} $ and integrating over the direction of $\vv$
one obtains a set of coupled equations for the moments of $W(x,v)$.
From now on we will use the temporal axial gauge $A_0 = 0$ which
is the most convenient for our purpose.
The zeroth moment of \eq{boltzmann} is the equation for current
conservation \mref{nonconservation}, \mref{intxi0}. Taking the first
moment of Eq.~\mref{boltzmann} gives
\begin{eqnarray}
        \partial_0 W^i(x) - [D^j,W^{ij}(x)] = \frac13 E^i(x) + \xi_0^i(x) 
        - \frac{N g^2 T}{4\pi}  \log(1/g) W^i(x) .
        \mlabel{wi}
\end{eqnarray}
The \lhs of Eq.~\mref{wi} is logarithmically
suppressed\footnote{Provided that $\partial_0\lsi g^2 T$, see below.}
relative to the term $\propto W^i(x)$ on the \rhs and can be
neglected. In this approximation we can determine $W^i(x)$ in terms of
$\vE(x)$ and $\xi_0^i(x)$ without solving any
differential equation. Introducing
\begin{eqnarray}
        \zeta^i(x) \equiv 4\pi \frac{\mmdebye}{Ng^2T}
        \frac{1}{\log(1/g)} \xi_0^i(x)
\end{eqnarray}
and 
\begin{eqnarray}
        \gamma = \frac{4\pi}{3}\frac{\mmdebye}{Ng^2 T}\frac{1}{\log(1/g)},
\end{eqnarray}
the spatial components of 
\eq{maxwellsoft} become
\begin{eqnarray}
        -\partial_0 E^i + [D_j,F^{j i}(x)] = 
        \gamma E^i(x) + \zeta^i(x) .
        \mlabel{langevin1}
\end{eqnarray}
The 0 component, or Gauss' law, now reads\footnote{For the
charge density in \eqs{gauss}-\mref{density} we use the same symbol as
for the one in \eqs{maxwell}-\mref{current}.The latter has both
soft and semi-hard Fourier components.}
\begin{eqnarray}
        [D_i,E^i(x)] = j_0(x) .
        \mlabel{gauss}
\end{eqnarray}
The charge density $j_0(x)=\mmdebye W(x)$ satisfies
\begin{eqnarray}
        \partial_0 j_0(x) = - \gamma j_0(x) - [D_i,\zeta^i(x)] .
        \mlabel{density}
\end{eqnarray}
Eqs.~\mref{langevin1}-\mref{density} form a closed
set of equations for the gauge fields and the charge density $j_0(x)$.
The stochastic force $\zeta^i(x)$ (cf.\ Eq.~\mref{xi0correlator4}) satisfies 
\begin{eqnarray}
        \Big\langle 
    \zeta^{ia}(x_1)
    \zeta^{jb}(x_{2}) 
    \Big\rangle 
        =
    2 T \gamma
        \delta^{ij } \delta^{ab}  \delta^{(4)}(x_1 - x_2).
        \mlabel{langevin2}
\end{eqnarray}

Eqs.~\mref{langevin1}-\mref{langevin2} are gauge covariant
Langevin-type equations. The dynamics of the soft fields is
Landau-damped by the hard field modes (see, e.g., Ref.\ \cite{landau10}).  The
kinematics of the hard modes being massless particles moving on
straight lines no longer appears in these equations because the
semi-hard fields randomize\footnote{I thank Guy Moore for this
  explanation.} the color of the hard particles on a length scale
$(\log(1/g) g^2 T)^{-1}$ which is small compared to the length scale
$(g^2 T)^{-1}$ of non-perturbative physics.  The random force
$\zeta^{ia}(x)$ is due to the thermal fluctuations of the initial
conditions for the semi-hard gauge field modes and for $w(x,v)$.

We will now solve the linearized form of the equation of motion
\mref{langevin1}. Even though we are interested in the
non-perturbative dynamics of the soft gauge fields, this will allow us
to determine the frequency scale on which the dynamics of these fields
becomes non-perturbative.  The solution for the transverse field reads
\begin{eqnarray}
        A_{\rm t}^{ia} (P) = -\frac{i}{p_0} p^2 
        \Big[ \bar{G}_{\rm t} (P) - \bar{G}_{\rm t} (0,\vp)\Big]
         A_{\rm t,in}^{ia} (\vp)
        + \bar{G}_{\rm t} (P) \[- E_{\rm t,in}^{ia} (\vp)
                + \zeta_{\rm t}^{ia} (P)  \] ,
\end{eqnarray}
where the subscript ``in'' refers to the  the initial values at $t=0$.
The transverse propagator $\bar{G}_{\rm t}$ is given by
\begin{eqnarray}
        \bar{G}_{\rm t} (P) \equiv \frac{1}{-P^2 - i\gamma p_0} .
        \mlabel{softpropagator}
\end{eqnarray}
The typical size of the soft field is $\vA_\tr (x)\sim \vA_{\rm t,in}
(\vx)\sim gT$. Thus both terms in a covariant derivative
$\partial_\mu - ig A_\mu$ are of the same size making the thermodynamics 
of the soft fields non-perturbative.

The dynamics of the soft gauge field becomes non-perturbative if it
changes in time by an amount $\Delta A$ which is of the same size as
$A$ itself. Then the non-linear terms in the equations of motion are
as important as the linear ones. $\Delta \vA_\tr$ can be estimated as
$\Delta \vA_{\rm t} (\vp) \sim p^2 \bar{G}_{\rm t} (P) \vA_{\rm t,in}
(\vp)$. From \eq{softpropagator} one reads off that $\Delta \vA_{\rm
  t} (\vp) \sim \vA_{\rm t,in} (\vp)$ when $p_0 \sim g^4 \log(1/g) T$.
Thus we find that large non-perturbative changes $\Delta \vA_\tr$ of
the transverse (magnetic) gauge fields are associated with the
frequency scale
\begin{eqnarray}
        p_0 \sim g^4 \log(1/g) T  . \mlabel{p0}
\end{eqnarray}

Let us now use this result to estimate the rate for electroweak baryon
number violation at very high
temperatures \cite{rubakov}, \cite{mclerran}-\cite{asy}, i.e., well above
the critical temperature $T_c\sim 100 $ GeV of the electroweak phase
transition or crossover. At such high temperatures the Higgs field
acquires a large thermal mass and decouples from the dynamics. Then it
is sufficient to consider a pure SU(2) gauge theory.  Baryon number
nonconserving processes are due to topology changing transitions in
the electroweak theory which are characterized by a change $\Delta
N_{\rm CS} = \pm 1$ of the Chern--Simons number
\begin{eqnarray}
   N_{\rm CS} = \frac {g^2}{32\pi^2} \epsilon_{ijk} \int d^3x
  \left( F^a_{ij} A^a_k - \frac{g}{3} \epsilon^{abc} 
  A^a_i A^b_j A^c_k \right) .
  \mlabel{chern}
\end{eqnarray}
Here $A_{i}^a$ are now the SU(2) gauge fields and $g$ is the weak
coupling constant. The change of baryon number $\Delta B$ is related
to $\Delta N_{\rm CS} $ by
\begin{eqnarray}
        \Delta B  = n_f \Delta N_{\rm CS} 
\end{eqnarray}
where $n_f$ is the number of fermion families. A change $\Delta N_{\rm
  CS} \sim 1$ requires the formation of a magnetic field configuration
with energy of order $(g^2 \Delta R)^{-1}$ where $\Delta R$ is its
spatial extent. In order for this configuration not to be Boltzmann
suppressed, its energy should not be larger than the temperature which
requires $\Delta R \gsim (g^2 T)^{-1}$. On the other hand, the size of
the field configuration cannot exceed the correlation length which is
of order $(g^2 T)^{-1}$. Thus the length scale relevant to the problem
is just the soft scale $(g^2 T)^{-1}$ at which finite temperature
perturbation theory breaks down. The amplitude of this field
configuration is $A\sim gT$.  As we have argued above, field
configuration of this size evolve with a frequency of order $ g^4
\log(1/g) T$ corresponding to a time scale $\Delta t\sim (g^4
\log(1/g) T)^{-1}$.  Thus the rate for a change of baryon number $B$
per unit time and unit volume can be estimated as
\begin{eqnarray}
\Gamma = \kappa g^2 \log(1/g) (g^2
T)^4 , \mlabel{rate}
\end{eqnarray}
where $\kappa$ is a non-perturbative coefficient which does not depend
on the gauge coupling~$g$. 

What might be the use of the effective theory for the soft gauge
fields derived in this letter for non-perturbative lattice computations
of real time correlators like \mref{c}?
The time scale for non-perturbative dynamics is much larger than the
corresponding length scale. Therefore the time derivative on the \lhs
of \eq{langevin1} can be neglected and one can write
\begin{eqnarray}
        [D_j,F^{j i}(x)] = 
        -\gamma \dot{A}^i(x) + \zeta^i(x) . 
        \mlabel{langevintag}
\end{eqnarray}
This equation should be easy to implement in a lattice calculation.
Note that we have obtained this equation by assuming that the
ultraviolet cutoff $\mu$ for the soft fields is as small as $\log(1/g) g^2
T$. Therefore the lattice cutoff should not be chosen too large. It
would be interesting to see whether the results obtained via
\eq{langevintag} depend on the cutoff. If they do, there might be a
way to include counter-terms such that the cutoff dependence is
canceled. 

It is conceivable that, unlike the hard thermal loops
\cite{bodeker95,arnold}, the effective theory for the soft fields in a
classical lattice gauge theory has a similar structure as in the
quantum theory. If this turned out to be the case, one could match the
coefficients in the classical counterpart of \eqs{langevin1} with the
coefficients obtained in this letter. Then a numerical computation in
the classical lattice theory should give the correct leading order
result for a non-perturbative correlation function. The difficulty of
this approach is that sub-leading contributions to the effective
theory are only logarithmically suppressed and can therefore not
easily be identified in numerical data. It might be possible to
improve the matching such that also non-logarithmic terms are
included. This would require an extension of the calculation presented
here.

In Ref.~\cite{mooreparticles} an algorithm was employed which, in
addition to classical gauge fields, contains massless particles which
interact weakly with the gauge fields. This approach solves the
problem of implementing hard thermal loops on a lattice and it appears
to be equivalent to the starting point for the present calculation,
i.e., the hard thermal loop effective theory described by
\eqs{maxwell}-\mref{hamiltonian}.  With this algorithm it is again
difficult to distinguish logarithmic from non-logarithmic
contributions. Probably only a combination of the different methods
discussed above can determine the rate for electroweak baryon number
violation.

{\bf Acknowledgments.} I am grateful to E.~Berger, W.~Buchm\"uller,
A.~Jakovac, K.~Kainulainen, K.~Kajantie, A.~Kovner, M.~Laine, G.~D.~Moore,
H.~Nachbagauer, O.~Nachtmann, O.~Philipsen, T.~Prokopec, A.~Rebhan,
M.G.~Schmidt and C.~Wetterich for useful discussions. 
This work was supported in part by the TMR network
``Finite temperature phase transitions in particle physics'', EU
contract no. ERBFMRXCT97-0122.

\end{document}